\documentclass[prl,preprint,preprintnumbers,amsmath,amssymb]{revtex4}

\usepackage{graphicx}
\usepackage{graphics}
\usepackage{dcolumn}
\usepackage{bm}
\usepackage{amssymb}
\usepackage{wasysym}
\usepackage{sidecap}
\usepackage{bibtopic}

\begin{document}

\def\Ef{$E_{\rm F}$}
\def\Eb{$E_{\rm B}$}
\def\Efmath{E_{\rm F}}
\def\Ed{$E_{\rm D}$}
\def\Tc{$T_{\rm C}$}
\def\kpara{{\bf k}$_\parallel$}
\def\kparamath{{\bf k}_\parallel}
\def\kperp{{\bf k}$_\perp$}
\def\Gbar{$\overline{\Gamma}$}
\def\Kbar{$\overline{K}$}
\def\Mbar{$\overline{M}$}
\def\BiTe{Bi$_2$Te$_3$}
\def\BiSe{Bi$_2$Se$_3$}
\def\Ed{$E_{\rm D}$}
\def\invA{\AA$^{-1}$}

\title {
Photoemission of \BiSe\ with Circularly Polarized Light: \\
Probe of Spin Polarization or Means for Spin Manipulation?}

\author{J. S\'anchez-Barriga$^1$,  A. Varykhalov$^1$, J. Braun$^2$, S.-Y. Xu$^3$, N. Alidoust$^3$, O. Kornilov$^4$, J. Min\'ar$^2$, K. Hummer$^5$, G. Springholz$^6$, G. Bauer$^6$, R. Schumann$^4$, L. V. Yashina$^7$, H. Ebert$^2$, M. Z. Hasan$^3$ and O. Rader$^1$}

\affiliation{$^1$Helmholtz-Zentrum Berlin, Albert-Einstein-Str. 15, 12489 Berlin, Germany}
\affiliation{$^2$Department Chemie, Ludwig-Maximilians-Universit\"at M\"unchen, 
Butenandtstr. 5-13, 81377 M\"unchen, Germany}
\affiliation{$^3$Joseph Henry Laboratory and Department of Physics, Princeton University, Princeton, New Jersey 08544, USA}
\affiliation{$^4$Max-Born-Institut, Max-Born-Str. 2A, 12489 Berlin, Germany}
\affiliation{$^5$ University of Vienna, Faculty of Physics, Computational Materials Physics, Sensengasse 8/12, 1090 Vienna, Austria}
\affiliation{$^6$Institut f\"ur Halbleiter- und Festk\"orperphysik,
Johannes Kepler Universit\"at,  Altenbergerstr. 69, 4040 Linz, Austria}
\affiliation{$^7$Department of Chemistry, Moscow State University, Leninskie Gory 1/3, 119991, Moscow, Russia}

\begin{abstract}
{\bf Topological insulators are characterized by Dirac cone surface states with electron spins aligned in the surface plane and perpendicular to their momenta. Recent theoretical and experimental work implied that this specific spin texture should enable control of photoelectron spins by circularly polarized light. However, these reports questioned the so far accepted interpretation of spin-resolved photoelectron spectroscopy. We solve this puzzle and show that vacuum ultraviolet photons (50-70 eV) with linear or circular polarization probe indeed the initial state spin texture of \BiSe\ while circularly polarized 6 eV low energy photons flip the electron spins out of plane and reverse their spin polarization. Our photoemission calculations, considering the interplay between the varying probing depth, dipole selection rules and spin-dependent scattering effects involving initial and final states explain these findings, and reveal proper conditions for light-induced spin manipulation. This paves the way for future applications of topological insulators in opto-spintronic devices.}
\end{abstract}

\maketitle

Since the discovery of three-dimensional topological insulators (TI's), the spin properties of their surface states have been of central importance to the field \cite{Hasan-RMP-2010}. Spin-resolved angle-resolved photoemission spectroscopy (SR-ARPES) has become the most powerful and the sole tool in systematically revealing the spin polarization of the topological surface states (TSS's) in energy and momentum space. Understanding and utilization of the spin properties of TI materials are believed to be the key of measuring the topological invariances hidden in the bulk electronic wave functions \cite{KanePRL2005}, realizing exotic magnetic-spin physics such as the axion electrodynamics \cite{Essin, Zhang} or the magnetic monopole \cite{Monopole}, as well as future spin-based low power transistors and devices. While extensive SR-ARPES studies on various TI compounds have successfully identified and confirmed the helical spin texture of the TSS's \cite{DilReview, Hsieh-Science-Nature-09, Souma11, Pan11, Jozwiak11}, much remains to be addressed regarding the critical response of the measured spin properties to the incident light and its polarizations (electric or magnetic fields).

In spite of the considerable success of SR-ARPES, it has been recently challenged by other experimental \cite{Wang-PRL-2011, Baharamy-arxiv-2012} and theoretical \cite{Henk, Louie-PRL-2012} proposals regarding both the efficiency and the reliability of the SR-ARPES measurements in studying the spin properties of TI surfaces. Part of these proposals involve the interpretation of the circular dichroism in the angular distribution (CDAD) of ARPES as the spin polarization of TSS's \cite{Wang-PRL-2011, Baharamy-arxiv-2012, Henk}, which, if correct, can significantly improve the efficiency of spin detection. The CDAD effect has also been predicted as an indirect measure of the intrinsic momentum-space orbital angular momentum texture of the TSS's \cite{Jung-PRB-2011, Park-PRL-2012}. These interpretations are, however, unrealistic for several reasons and one is the dominance of final-state effects in CDAD from TI's which changes sign several times with photon energy \cite{ScholzPRL13}. Because of this dominance of final-state effects in an ARPES-based method, the influence of the photoemission process and the significance of conclusions drawn from SR-ARPES of TI's, so far conducted using linearly polarized light, are under question. 

On the other hand, in the presence of strong spin-orbit coupling, such predicted orbital texture would likely be better coupled to and thus easier controlled by electric fields which can be utilized in coherent spin rotation \cite{Nowack}, spin-orbit qubits \cite{Kouwenhoven}, as well as photon-polarization driven spin current devices based on the spin-orbit coupled electrons on TI surfaces \cite{Gedik_NT}. In this context, a recent theoretical work \cite{Louie-PRL-2012} has also raised concerns regarding the reliability of SR-ARPES in properly revealing the spin polarization of TSS's, in particular concerning the response of the measured spin properties to the polarization of the incident light. Assuming strong spin-flip or spin-rotation effects during the photoemission process, it has been proposed that the spin texture of the photoelectrons (i.e., of the final states) is completely different from that of the TSS's in the initial state when linearly or circularly polarized light is used under specific experimental conditions and sample geometries, depending on the angle between light polarization and initial state spins \cite{Louie-PRL-2012}. Consequently, the measured spin texture of the photoelectrons should completely "lose memory" of the initial state spin texture and instead rotate depending on the chosen polarization of the incident light. However, it is important to note that the previously existing SR-ARPES data have been interpreted under the assumption that electron spins emitted from TI surfaces are conserved in the photoemission process. Following the first theoretical work \cite{Louie-PRL-2012}, a recent SR-ARPES experiment using laser light of 6 eV photon energy \cite{JozwiakNP13} has reported spin-resolved data strongly supporting the spin-rotation final-state scenario, and it was concluded that this scenario essentially dominates the spin polarization of photoelectrons emitted from TI's \cite{JozwiakNP13, XueNP13}. While such final-state effects are of high interest for the purpose of spin manipulation in opto-spintronics applications, these theoretical and experimental studies do on the other hand also strongly challenge the reliability and robustness of SR-ARPES in studying the spin properties of TI surfaces. Considering the fact that SR-ARPES is presently the only available tool for such a purpose, it is critically important to systematically study the impact of such final-state effects in SR-ARPES measurements under different experimental conditions. Namely, under which conditions the proposed final state effects are merely a weak perturbation on the initial spin texture and when they become enhanced or even dominate as proposed recently \cite{Louie-PRL-2012, JozwiakNP13, XueNP13}.

In the present work, we utilize the TSS of the prototype TI \BiSe\ as a platform to systematically investigate the recently proposed basis of manipulating the spin orientation of photoelectrons emitted from the surface of TI's by the polarization of the incident photons \cite{Louie-PRL-2012}. By using different experimental conditions and sample geometries, we investigate the reliability of SR-ARPES in studying the initial state spin texture of TSS's with linearly and circularly-polarized photons. We demonstrate the existence of two limiting cases where spin manipulation with light polarization is either a weak perturbation or strongly dominates the photoemission process. We further identify the underlying mechanism that triggers the spin polarization and support our experiments by comparing to results of one-step model photoemission calculations.\\

\noindent {\bf RESULTS}

We perform SR-ARPES measurements on \BiSe\ films with linear p- and circularly polarized light of opposite helicities (C+ and C-) incident at an angle of $\phi=45^{\circ}$ with respect to the sample normal. Because it is theoretically expected that light-induced manipulation of photoelectron spins must depend on the angle between light polarization and initial state spin \cite{Louie-PRL-2012}, we use two different sample geometries shown in Fig. 1(a) (geometry I) and Fig. 2(a) (geometry II) [for more details please see Methods section and Supplemental Material \cite{Supplement}]. Figure 1 shows SR-ARPES results of the in-plane spin polarization of the TSS and bulk valence band states of \BiSe\ measured using sample geometry I with different photon energies and light polarizations. In order to investigate the predicted spin-rotation final-state effect \cite{Louie-PRL-2012}, we first reverse the helicity of the circular light polarization to search for photon energies where the circular dichroism in ARPES changes its sign. We find for \BiSe\ pronounced sign changes in the CDAD signal between 50 and 70 eV. We note that these are different from the ones reported for \BiTe\ \cite{ScholzPRL13}. Specifically, the color representation of the CDAD asymmetry $A$=$\frac{I(C+)-I(C-)}{I(C+)+I(C-)}$ (where $I$ is the photoemission intensity) shows that for the upper Dirac cone it reverses from 50 eV [Fig. 1(b)] to 60 eV [Fig. 1(e)], while for the lower Dirac cone from 60 eV to 70 eV [Fig. 1(h)]. Therefore, we search at these photon energies for coupling to final-state effects in the {\it spin polarization} caused by the circular light polarization.

We select a few \kpara\ wave vectors which are marked by green dashed lines in Figs. 1(b), 1(e) and 1(h). Upon reversal of the circular light polarization, changes (for fixed \kpara\ and photon energy) are neither observed in the spin-resolved spectra [Figs. 1(c1)-1(c2), 1(f1)-1(f2) and 1(i1)-1(i2)] nor in the corresponding spin polarizations [Figs. 1(d1)-1(d2), 1(g1)-1(g2) and 1(j1)-1(j2)]. In addition, the spectra are indistinguishable from measurements with linear light polarization [symbol $\leftrightarrow$ in Figs. 1(c3)-1(c4)]. The out-of-plane spin polarization under this sample geometry was found to contribute unobservably to the measured spin polarizations (not shown). The fact that the in-plane spin polarization does not change indicates that the circularly polarized light does not align the spins parallel to the surface normal for C- and antiparallel for C+ polarization, in spite of what has been predicted \cite{Louie-PRL-2012}. According to Ref. \onlinecite{Louie-PRL-2012}, either zero or very small in-plane spin polarizations would have been expected for C- as well as C+ in Fig. 1. 

We further demonstrate in Fig. 2 that a weak coupling to the light polarization indeed exists by showing SR-ARPES results of the in-plane as well as the out-of-plane spin polarization measured at 50 eV using sample geometry II. The spin perpendicular to the surface does not reach the predicted 100\%\ but stays small ($\sim10$\%). This small polarization is not reversed between C+ [Fig. 2(d4)] and C- [Fig. 2(h2)], and does also not change when switching from circular to linear light polarization [Fig. 2(f2)]. Instead, the out-of-plane spin polarization reverses with \kpara\ [Figs. 2(d2)], indicating that it is an initial-state effect. The reason is hexagonal warping \cite{Fu-PRL-2009} near the Fermi surface [see Fig. 2(i) (top)] which is rather small in \BiSe\ as compared to \BiTe\ \cite{Souma11}. 

Moreover, for {\it linearly} polarized light, the angle between linear light polarization and spin should lead to a change of the spin in the final state  \cite{Louie-PRL-2012}. This results in a complex in-plane spin texture where the circulation of photoelectron spins around the Fermi surface appears similar to the one followed by a magnetic field in an idealized quadrupole \cite{Louie-PRL-2012, JozwiakNP13}. Light polarization and in-plane spins are parallel in geometry II and perpendicular in geometry I but Fig. 2(f1) shows that the in-plane spin polarization has not changed compared to Fig. 1(d4), demonstrating that it is indeed independent of the sample geometry and follows the expected \cite{Hsieh-Science-Nature-09} initial state helical texture of surface-state electron spins.

The maximum observed in-plane spin polarization in Figs. 1-2 is about $(50 \pm 10)$\%. This deviates from recent reports of nearly 100\% spin polarization from TSS's in \BiSe\ \cite{Pan11, Jozwiak11}. We do not observe large nonzero spin polarization values at the Dirac point \cite{Jozwiak11} or {\bf k}-independent spin polarizations contributing to the specific spin texture of TSS's in the photoelectron distribution \cite{Jozwiak11}. Considering the multiple orbital origin of the \BiSe\ surface states and their multiple contribution to the net spin polarization, a magnitude of $\sim$50\%, as reported here, is consistent with first-principles calculations of \BiSe\ \cite{Yazyev-PRL-2010}. We do note that in Fig. 1 the measured spin polarization magnitude of the lower Dirac cone at photon energies of 50 eV and 60 eV appears considerably reduced as compared to that at 70 eV. This can be understood based on the superposition of photoemission from the TSS and the bulk valence band, as the bulk bands in \BiSe\ are essentially unpolarized. Hence, at photon energies (effectively \kperp\ values) where the contribution of the bulk valence band is weak (such as 70 eV), we find that the spin polarization measured from the lower Dirac cone accordingly increases up to the expected value \cite{Yazyev-PRL-2010}. 

Figure 2(i) summarizes our experimental findings. The handed helical spin texture of the TSS (bottom), which reverses at the Dirac point, is observed along four \kpara\ directions ($\pm \overline{\Gamma}\overline{K}$ and $\pm \overline{\Gamma}\overline{M}$) (top). The corresponding photoelectron spin directions in-plane (green arrows) and out-of-plane (pink arrows) remain in orientation and magnitude unchanged for the different light polarizations (C+, C-, $\leftrightarrow$). This does not change over an extended range of incident photon energies (50--70 eV), although the ARPES spectral weight distribution and CDAD signals have changed drastically within the same photon energy range. Thus, we conclude that at these photon energies, both the direction and magnitude of the measured photoelectron spin polarization remains representative of the initial TSS's and that the theoretically expected photoelectron spin reorientation contributes very little and almost unobservably to the measured spin polarizations. 

However, we find a completely different situation when using circularly polarized light at 6 eV photon energy, in agreement with recent findings \cite{JozwiakNP13}. Our laser SR-ARPES results presented in Fig. 3 reveal that circularly polarized photons flip the photoelectron spins perpendicular to the surface and reverse the resulting out-of-plane spin texture with the sense of circular polarization. Specifically, the in-plane component of the spin polarization is nearly zero [upper panels in Figs. 3(b)-3(g)] and the out-of-plane one as large as $\sim46$--$53$\% [bottom panels in Figs. 3(b)-3(g)], which is comparable to the polarization value of the in-plane component measured at 50--70 eV in Figs. 1-2. Moreover, the out-of-plane spin polarization does not reverse when going from $+$\kpara\ to $-$\kpara\ wave vectors [compare Figs. 3(c2) and 3(e2)], again differently from the behavior at 50--70 eV. This suggests that light-induced manipulation of the photoelectron spin must depend on the final states reached at different photon energies, a fact which was not considered in previous theoretical models where so far only spin-degenerate free-electron-like final states have been taken into account.\\   

\noindent {\bf DISCUSSION}

To resolve this issue and further explore the origin of these two contrasting experimental findings, we have performed one-step photoemission calculations including spin-dependent transition matrix elements between initial and final states. Figure 4(a) shows the in-plane and out-of-plane components of the photoelectron spin polarization calculated at 50 eV photon energy upon reversal of the light helicity using a light incidence angle of $\phi=45^{\circ}$. The binding energy is 0.16 eV, where the constant energy surface above the Dirac point is mainly of circular shape. The calculated in-plane photoelectron spin polarization of $\sim$63\% remains unaffected upon reversal of the circular photon polarization, thus reflecting the initial state spin texture of the TSS in good agreement with our experimental results in Figs. 1-2. Our calculations reveal that the absolute value of the spin polarization scales with the light incidence angle as $\sim|\cos\phi|$. Thus, similar calculations to the ones shown in Fig. 4(a) but in normal incidence lead to in-plane spin polarization values of up to $\sim$90\%. We also do note that in Fig. 4(a) the out-of-plane spin switching effect for opposite light helicities contributes with less than 4\% spin polarization in the background of the present calculations (the color representation, which is proportional to the magnitude of the spin polarization, has been magnified in the upper right quadrants corresponding to calculations of the out-of-plane spin polarization). On the other hand, the higher absolute theoretical values of the in-plane spin polarization as compared to the experiment are most probably due to a theoretical overestimation of the surface contribution (see below). 

The fact that in \BiSe\ free-electron final states are not accessible at photon energies of the order of 50--70 eV becomes clear from, e.g., the reversal of the CDAD effect in ARPES (see Fig. 1), which involves transitions from $p$-type initial states to $d$-type final states \cite{ScholzPRL13, Supplement}. To reach spin-degenerate free-electron final states in a reasonable approximation, much higher photon energies are required. Figure 4(b) shows calculations in normal incidence for such a condition (300 eV), where we find the result that the out-of-plane spin polarization reverses completely with the circular polarization, as theoretically predicted \cite{Louie-PRL-2012}. It is surprising that an experiment at 6 eV appears to confirm a theory that invokes free-electron final states. Therefore, we also performed calculations at 6 eV for comparison (please see Supplemental Material for details \cite{Supplement}). Our 6 eV calculations reveal that positive and negative circularly polarized photons reverse $\sim$25-30\% out-of-plane spin polarization, in qualitative agreement with the calculation in Fig. 4(b) and the experiments in Fig. 3, despite the somewhat smaller absolute values of the spin polarization. The origin of this similarity to the high-energy case is partially linked to dipole selection rules because 6 eV photon energy enables transitions into $s$-type final states which are also spin-degenerate. On the other hand, the smaller calculated values of the out-of-plane and the nearly zero values of the in-plane spin polarization at low photon energies are partially related to the effect of multiple scattering in the final states which is fully included in our calculations via spin-dependent scattering matrix elements \cite{Supplement}.

The above results show that we have to distinguish between three different spectral ranges, and this is also related to different inelastic mean free paths. The existence of light-induced spin polarization control depends not only strongly on the final states, as discussed above, but additionally on the probing depth of the photoelectrons, which is considered in our theoretical model. The TSS of \BiSe\ extends several layers deep into the bulk ($\sim$ 2-3 nm) and therefore appears for a wide range of probing depths. Indeed, enhanced bulk sensitivity at 6 eV photon energy (i. e., $\sim1$ eV kinetic energy) is provided by an order of magnitude larger probing depth than at 50--70 eV \cite{Koralek07}, and similarly at photon energies in the x-ray range. Specifically, our detailed analysis of the initial state identifies the low binding energy states in \BiSe\ as mainly arising from different $p$-orbitals of Bi (6p$^{3}$) and Se (4p$^{4}$). This results in a layer-dependent spin-orbital texture with a large out-of-plane $p_{z}$ orbital character within the first quintuple layer ($\sim$ 1 nm), and a significant in-plane $p_{x,y}$ contribution deeper below the surface. As a consequence, entanglement between orbital and spin-degrees of freedom due to strong spin-orbit coupling is present in our calculations when deeper lying layers are taken into account, as reported also in recent independent calculations \cite{Zhu-PRL-2013}. In combination with dipole selection rules for different final states, this causes the light-induced out-of-plane photoelectron spin polarization due to interference effects which can be followed in the calculated spin-dependent scattering matrix elements \cite{Supplement}. This further implies that light-induced control of photoelectron spins is less effective near the surface, the spin polarization of which is subject to the effect of multiple scattering.

To summarize, we have experimentally and theoretically investigated the recently proposed suggestion of manipulating the spin polarization of photoelectrons from TI's in SR-ARPES experiments, according to which it should be entirely controlled by the light polarizations. For 50-70 eV photons we have found experimental conditions for which the photoelectron spin polarization measured in SR-ARPES experiments using linear p- and circularly polarized light is the proper probe for the initial state spin texture of the TSS of \BiSe. We have shown for \BiSe\ that neither the in-plane ($\sim$50\%) nor the out-of-plane spin polarization ($\sim$10\%) changes when the polarization of this vacuum ultraviolet light is switched from linear to circular and positive to negative helicity, in full agreement with our one-step-photoemission calculations. However, we have shown that positive and negative circularly polarized 6 eV photons switch the spins perpendicular to the surface, offering manipulation of photoelectron spins with light polarization for this photon energy. Based on our one-step model calculations, we have demonstrated that these two contrasting experimental findings for the energy range from 50 to 70 eV on the one hand and for 6 eV on the other are due to the interplay between initial and final states, bulk sensitivity and dipole selection rules in the spin-dependent scattering matrix elements. Our results provide important implications in both understanding and potentially controlling spin degrees of freedom of emitted photoelectrons from TI surfaces using light and its polarizations.\\

\noindent {\bf Methods}

SR-ARPES experiments on \BiSe\ films are performed at room temperature in ultrahigh vacuum better than $1\cdot10^{-10}$ mbar. Experiments at photon energies between 50--70 eV are carried out with polarized undulator radiation at the UE112-PGM1 beamline of BESSY II. The 4th harmonic of a home-made fs-laser with central wavelength $\lambda$=800 nm and a pulse duration $\sim$150 fs is used to create circularly polarized light of 6 eV photon energy. Spin analysis of the photoelectrons is provided by a Rice University Mott-type spin polarimeter \cite{Burnett-RSI-94} coupled to a Scienta R8000 hemispherical analyzer. The energy resolution of the SR-ARPES experiment is $\sim$80 meV and the angular resolution $\sim$0.8$^{\circ}$ \cite{Supplement}. Sample geometries I and II are reached by keeping the rotation about the surface normal (azimuth angle) fixed and rotating either the polar $\theta$ or the tilt $\gamma$ angles of the sample, respectively. \BiSe\ films 400 nm thick are grown by molecular beam epitaxy on BaF$_2$(111) substrates using a \BiSe\ compound and elemental Se effusion cells. During deposition the substrate temperature is kept at $360\,^{\circ}{\rm C}$ under which condition two-dimensional growth is observed by {\it in situ} reflection high-energy electron diffraction. Calculations of the SR-ARPES intensity are based on multiple scattering theory within the one-step model of photoemission including wave-vector, spin and energy-dependent transition matrix elements \cite{Hop80, Bra96}. Initial and final states are obtained for a semi-infinite half-space using the low-energy electron diffraction method \cite{gray11}. Further theoretical and experimental details can be found in the Supplemental Material \cite{Supplement}.\\ 

\noindent {\bf Acknowledgements}

This work was supported by SPP 1666 of the Deutsche Forschungsgemeinschaft. S.-Y. X., N. A. and M. Z. H. are supported by the Office of Basic Energy Sciences, US Department of Energy (X-ray program DOE/BES Grant No. DE-FG-02-05ER46200).

\newpage

\noindent {\large\bf Figure Captions}\\

\noindent {\bf Figure 1.} 
Observation of the initial state in-plane spin polarization upon reversal of the CDAD effect from the TSS using (a) geometry I (where the polar angle $\theta$ of the sample is rotated), different light polarizations [linear ($\leftrightarrow$), circular positive (C+), circular negative (C-)] and photon energies. [(b),(e),(h)]: {\bf k}-resolved CDAD asymmetry shown for each photon energy. The green dashed lines on top mark the {\bf k}$_\parallel$ wave vectors for which SR-ARPES measurements are presented. The CDAD values from the TSS near the Fermi level are given at the top. [(c),(f),(i)]: Spin-resolved energy distrubution curves (EDC's) measured for different wave vectors, photon energies and light polarizations. The black (red) curves show tangential spin up (down) EDC's. [(d),(g),(j)]: To the right of each spin-resolved EDC's, the corresponding net tangential spin polarization is shown and its magnitude indicated.\\


\noindent {\bf Figure 2.} 
In-plane and out-of-plane spin polarizations from the TSS measured at 50 eV using (a) geometry II (where the tilt angle $\gamma$ of the sample is changed) as well as different light polarizations. (b) ARPES intensity obtained with linear p-polarized light. [(c),(e),(g)]: In-plane [(c1),(c3),(e1),(g1)] and out-of-plane [(c2),(c4),(e2),(g2)] spin-resolved EDC's measured for various {\bf k}$_\parallel$ wave vectors [marked by white dashed lines in (b)]. [(d),(f),(h)]: To the right of each spin-resolved EDC's, the corresponding net spin polarizations. (i) Summary of the results from SR-ARPES measurements at 50 eV; (Top) The handed helical spin texture is probed in four \kpara\ directions ($\pm \overline{\Gamma}\overline{K}$ and $\pm \overline{\Gamma}\overline{M}$). The Fermi surface exhibits a small distortion due to hexagonal warping. The in-plane (green arrows) and out-of-plane (pink arrows) photoelectron spin directions are in vector and magnitude independent of the light polarization. (Bottom) Schematics of the observed in-plane helical spin texture superimposed on constant energy surfaces extracted from the data shown at the top.\\


\noindent {\bf Figure 3.}
Out-of-plane switching of the spin texture from the TSS observed upon reversal of the circular light polarization at 6 eV photon energy. (a) {\bf k}-resolved CDAD asymmetry. The green dashed lines mark the {\bf k}$_\parallel$ wave vectors of the SR-ARPES measurements. In-plane [(b1),(d1),(f1)] and out-of-plane [(b2),(d2),(f2)] spin-resolved EDC's. [(c),(e),(g)]: Corresponding net spin polarizations. The out-of-plane spin polarization (zero in-plane) does not reverse when going from $+$\kpara\ to $-$\kpara\ wave vectors and switches with the light helicity. Spectra were acquired using geometry II as well as [(b)-(e)] circular negative (C-) and [(f),(g)] circular positive (C+) light polarizations.\\  


\noindent {\bf Figure 4.}
Calculations of the in-plane (left panels) and out-of-plane (right panels) spin polarization of photoelectrons emitted from the TSS upon reversal of the light helicity. (a) Results at 50 eV under 45$^\circ$ light incidence angle and (b) for free-electron-like final states (300 eV) in normal incidence. Arrows in (a) indicate the in-plane spin polarization directions, dots and crosses in (b) the spin polarization parallel and antiparallel to the surface normal, respectively. The color representation is proportional to the magnitude of the spin polarization and it has been magnified in the upper right quadrants of the calculations corresponding to the in-plane and out-of-plane spin polarization in (b) and (a), respectively.\\

\newpage
\centering
\includegraphics [width=1\textwidth]{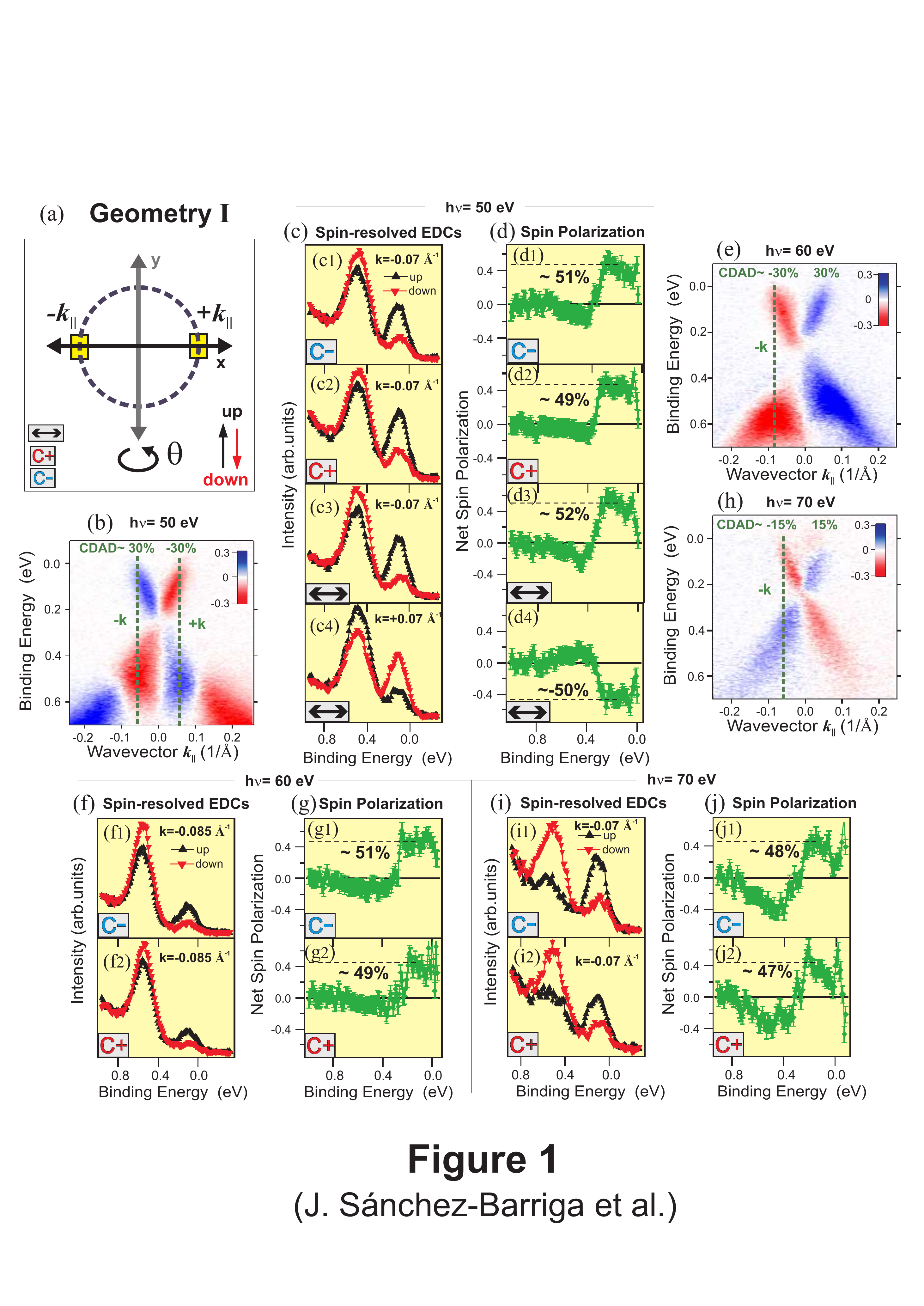}

\newpage
\centering
\includegraphics [width=1\textwidth]{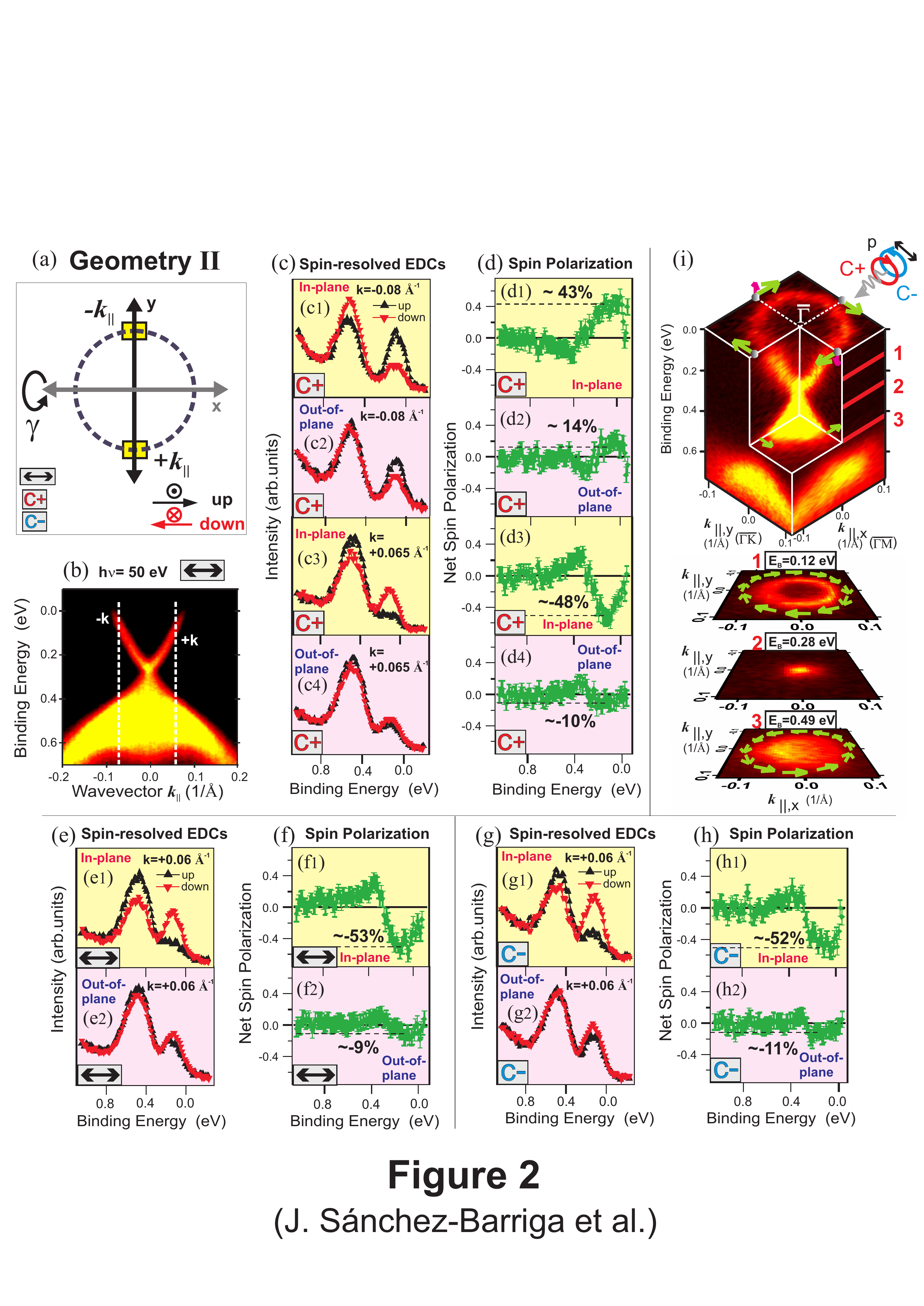}

\newpage
\centering
\includegraphics [width=1\textwidth]{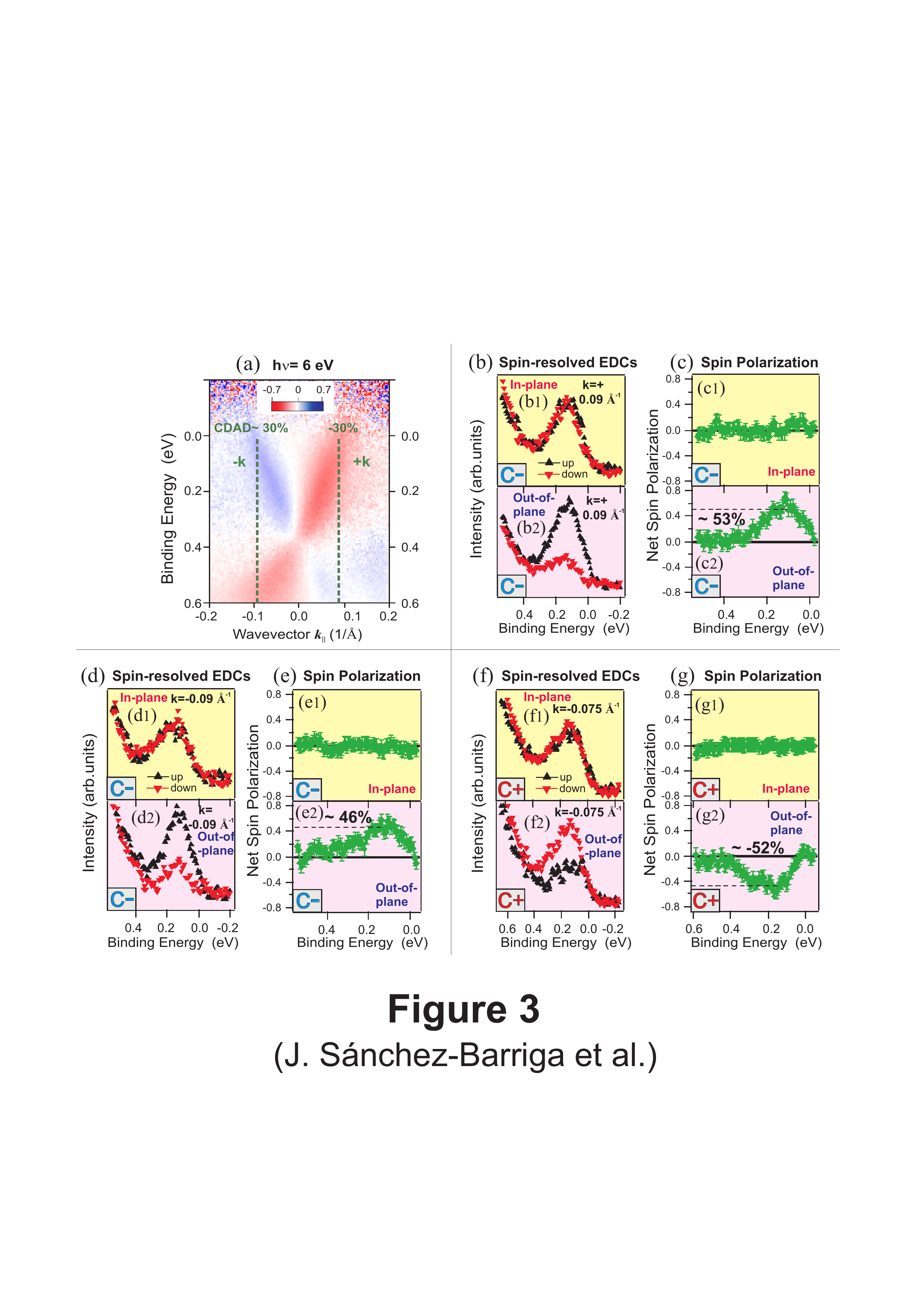}

\newpage
\centering
\includegraphics [width=1\textwidth]{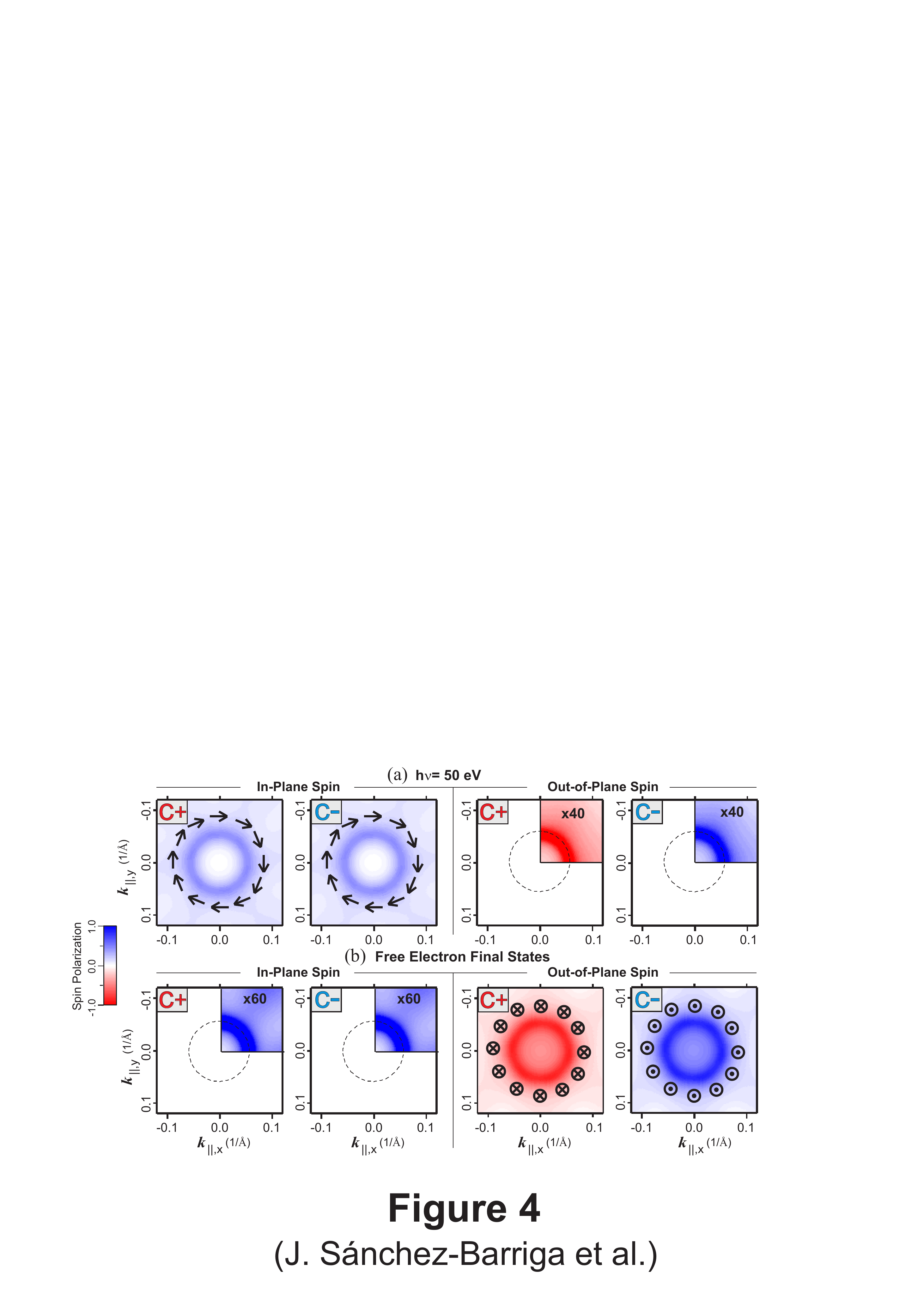}

\end{document}